\documentclass[aps,prb,showpacs,twocolumn]{revtex4-2}

\usepackage{amssymb}
\usepackage{graphicx}
\usepackage{amsmath,amssymb,color,bm}
\usepackage{wrapfig}

\begin{document}
	
	\title{Contribution to the GdTe$_3$ magnetic phase diagram from electronic transport measurements}
	
	\author{A.A.~Sinchenko$^{1}$,  P.D. Grigoriev$^{2,3}$, A.~Hadj-Azzem$^{4}$, E.~Pachoud$^{4}$, E. Lorenzo$^{4}$ and P.~Monceau$^{4}$}
	
	\date{\today}

	\affiliation{$^{1}$Kotelnikov Institute of Radioengineering and Electronics of RAS, 125009 Moscow, Russia}
	\affiliation{$^{2}$L. D. Landau Institute for Theoretical Physics, 142432, Chernogolovka, Russia}
	\affiliation{$^{3}$National University of Science and Technology "MISiS", 119049 Moscow, Russia}
	\affiliation{$^{4}$Institut Néel, Université Grenoble Alpes, CNRS, 38000, Grenoble, France}
	
	\begin{abstract}
		
			Magnetic phase diagrams of the three antiferromagnetic (AFM) transitions of GdTe$_3$ that occur at zero magnetic field at $T_{N1}=12$ K, $T_{N2}=10.5$ K and $T_1=7$ K are investigated by means of resistance measurements in a magnetic field oriented along the three crystallographic directions. With the magnetic field applied in the basal plane we found re-entrance of an intermediate phase which may be a result of the coupling between the $T_{N2}$ AFM and the charge density wave with possibly a ferroaxial order. A spin flop transition is also observed at a temperature that presumably converges towards a triple point in vicinity of 2.5-3.0 T and $T$ between $T_1$ and $T_{N2}$. Our results help complete a complex phase diagram of GdTe$_3$ and advance a method for investigating magnetic ordering via transport measurements -- a technique applicable to many other compounds.
		
	\end{abstract}

	\maketitle
	
	\section{introduction\label{INTR}}
	
The interplay and coexistence of several ground states at low temperature between charge density wave (CDW) and different types of magnetic ordering have recently attracted considerable interest in condensed matter. This interplay may be realized in low-dimensional metallic compounds containing rare-earth elements, where the crystal lattice, conduction electrons and 4f magnetic moments interact, as intensively studied in various systems such as divalent europium compounds EuX$_4$ (X=Al, Ga)\cite{Kaneko21}, rare-earth intermetallic \textit{R}NiC$_2$ (\textit{R}=Gd, Tb) \cite{Shimomura16} and Er$_5$Ir$_4$Si$_{10}$ \cite{Galli2000}, and van der Waals rare-earth tritellurides RTe$_3$ (R=La, Ce, Pr, Nd, Sm, Gd, Tb, Dy Ho, Er, Tm) \cite{Yumigeta20}.

All the RTe$_3$ compounds have the same orthorhombic structure (\textit{Cmcm}) in the unmodulated state. The structure is formed by blocks of [R$_2^{3+}$Te$_2^{2-}$] layers sandwiched between two Te layers and stacked together along the $b$-axis with a weak van der Waals gap ($\sim 3.8$ \AA) between the Te layers \cite{remark1}. In-plane $p_x$ and $p_z$ orbitals of Te atoms from square nets dominate electronic states near $E_F$. Three electrons are donated per R$^{3+}$ ion in RTe$_3$, two of these electrons fill the Te$_{5p}$ bands of RTe slabs, while one electron transfers to the Te square-net sheets. This results in the RTe slabs being insulating and the conducting electrons well confined within Te square-net sheets. The band structure calculated with a simple tight-binding approximation reveals that the electronic bands at the Fermi level ($E_F$) come from the $p_x$ and $p_z$ in-plane Te orbitals, leading to a simple Fermi surface (FS). These systems exhibit an incommensurate CDW throughout the whole R series with a wave vector $\textbf{q}_{CDW1}=(0,0,\sim2/7c^*)$ with a CDW transition temperature above 300 K for the light atoms (La, Ce, Nd, Sm, Gd, Tb). For the heavier R (Dy, Ho, Er, Tm) a second CDW occurs with a wave vector $\textbf{q}_{CDW2}=(\sim2/7a^*,0,0)$. The superlattice peaks measured from x-ray diffraction are very sharp and indicate a long-range three dimensional (3D) CDW order.

The conventional CDW description assumes an equal modulation of the $p_x$ and $p_z$ orbital occupation, resulting in an $s$-type orbital condensation characterized by an order parameter with a scalar amplitude-excitation (Higgs) mode. In the case of multi component CDW with unequal orbital modulation of the $p_x$($p_z$) orbital occupation it has been recently shown \cite{Singh2025,Wang2022} that the CDW gives rise to an axial Higgs mode suggesting additional symmetry breaking which may point to the existence of hidden order. It has been proposed that this hidden order takes the form of coupled charge-and-orbital texture, leading to electronically driven ferroaxial order together with the opening of an axial electronic gap. This interpretation was challenged \cite{Freitas2026} by showing that a single component of the CDW order parameter with its wave vector away from the crystallographic $c$-axis reproduces the Raman spectra. Recent Raman studies have provided further evidence for the ferroaxial nature of the CDW in the RTe$_3$ family \cite{Singh2025,Wang2022}.

The physical separation between the localized and itinerant electrons can lead to multiple magnetic ground states resulting from the interplay between charge order and magnetic order in these RTe$_3$ materials. Indeed, thermodynamic (specific heat and magnetization) measurements show magnetic ordering of rare earth ions in RTe$_3$ at temperatures below 12 K. These temperatures are much lower than the CDW transition temperatures, and can, in general, be thought of as independent of the CDW ordering. However, resonant X-ray diffraction (RXD) measurements on TbTe$_3$, DyTe$_3$ and CeTe$_3$ have revealed a subtle interplay between the rare-earth 4f electrons with the 5p electrons of Te\cite{Lee2012}. Recent studies reported observation of a ground state with strongly coupled magnetic and CDW orders mediated most probably via orbital ordering. It has been shown \cite{Okuma2026} that another compound of the RTe$_3$ series, namely CeTe$_3$, hosts a correlated magnetic ground state where spin and itinerant charge degrees of freedom are intimately linked in the van der Waals limit. A theoretical study \cite{Aoyama2025} has considered that for the case when the CDW with an incommensurate ordering vector $q_{CDW}$ is induced by Fermi-surface nesting, the CDW order suppresses the conventional RKKY interaction and that can induce a mode coupling between magnetic ordering vectors $q_m$ and $q_m\pm q_{CDW}$. This suggests that, below the CDW transition temperature, the correlation between localized spins may develop at both $q_m$ and $q_m\pm q_{CDW}$. In fact, different types of magnetic peaks are observed from neutron scattering in TbTe$_3$ \cite{Pfuner2012,Chillal2020} and DyTe$_3$ \cite{Alekseev2024}: a commensurate reflection with $q_{AFM}=0.5c^*$ and an incommensurate CDW modulated AFM reflection with the propagation vector $q_{AFM}-q_{CDW}$.
	
In the series of $R$Te$_3$, the gadolinium tritelluride has raised special interest as it has the highest mobility \cite{Lei2020} of any van der Waals layered magnet, opening the possibility of its applications in low-temperature spintronics devices. Recent ARPES measurements show that this high carrier mobility can be attributed to the sharp and nearly linear band dispersions near the Fermi energy \cite{Liu2020}. GdTe$_3$ exhibits an incommensurate CDW transition at 380 K in the $(a,c)$-plane\cite{Ru2008} and a superconducting transition under pressure below 0.3 K \cite{Zocco15}. Using vertical stacking of strain-inducing layers, the formation of a one-dimensional (1D) moir\'{e} superstructure has been demonstrated in GdTe$_3$ \cite{Yeon25}. 

A cascade of low temperature magnetic transitions in GdTe$_3$ has been observed in the specific heat and magnetic susceptibility in the temperature range between 7 K and 12 K \cite{Lei2020,Iyeiri03,Ru08,Guo2021,Raghavan2024,Lei2021}, suggesting that the system passes through a series of magnetic ordered phases.  Two antiferromagnetic transitions at $T_{N1} = 11.5$ K and $T_{N2} = 9.7$ K and an anomaly at $T_1 = 7$ K were confirmed in $\chi(T)$ under a magnetic field of $\mu_0H = 1$ T \cite{Guo2021}. The magnetic susceptibility $\chi_{ac}$ drops more with decrease of $T$ than $\chi_{b}$ suggesting that Gd moment orders mainly in the basal plane. The anomaly at 7 K in $\chi(T)$ shows complicated changes in stronger magnetic fields \cite{Guo2021}. It was suggested that this anomaly around 7 K found in $\chi(T)$ and $C(T)$ might originate from the incommensurate CDW state in the ($a,c$)-plane of GdTe$_3$.  

Spin-polarized scanning tunneling microscopy directly images the charge and magnetic orders in GdTe$_3$\cite{Raghavan2024}. Two distinct magnetic orders -- an emergent spin density wave (SDW) with wave vector exactly equal to the CDW wave vector that sets in at $T\sim12$ K, and a striped antiferromagnetic order with wave vector \textbf{q$_m$}$=(\pm\pi/a,0)$ that sets in at $T\sim7.5$ K -- were identified. The authors proposed that a SDW occurs as a daughter order of an intermediate temperature $c$-axis AFM phase and the CDW, while the striped AFM is the actual low temperature ground state.

The exact nature of the magnetic phases in GdTe$_3$, especially for the 7 K anomaly, remains unknown. To make things worst, the very large absorption
cross-section of Gd precludes the use of neutron diffraction. It is, therefore, important to characterize the magnetic orders in order to unravel the unconventional mechanisms that are present in GdTe3. Under such conditions any new information about magnetic phases in GdTe$_3$ is important. In this regard, we present the results of electron transport measurements performed under various magnetic field configurations, with current applied along the $a$ and $c$ axes within the basal plane.

\section{experimental\label{EXP}}

Single crystals of GdTe$_3$ were grown by a self-flux technique under purified argon atmosphere as described previously \cite{SinchPRB12}. Thin single-crystal samples with a thickness less than 1 $\mu$m were prepared by micromechanical exfoliation of relatively thick crystals glued on a sapphire substrate. The quality of selected crystals and the spatial arrangement of the crystallographic axes were controlled using X-ray diffraction techniques. 

\begin{figure}[t]
	\includegraphics[width=8.5cm]{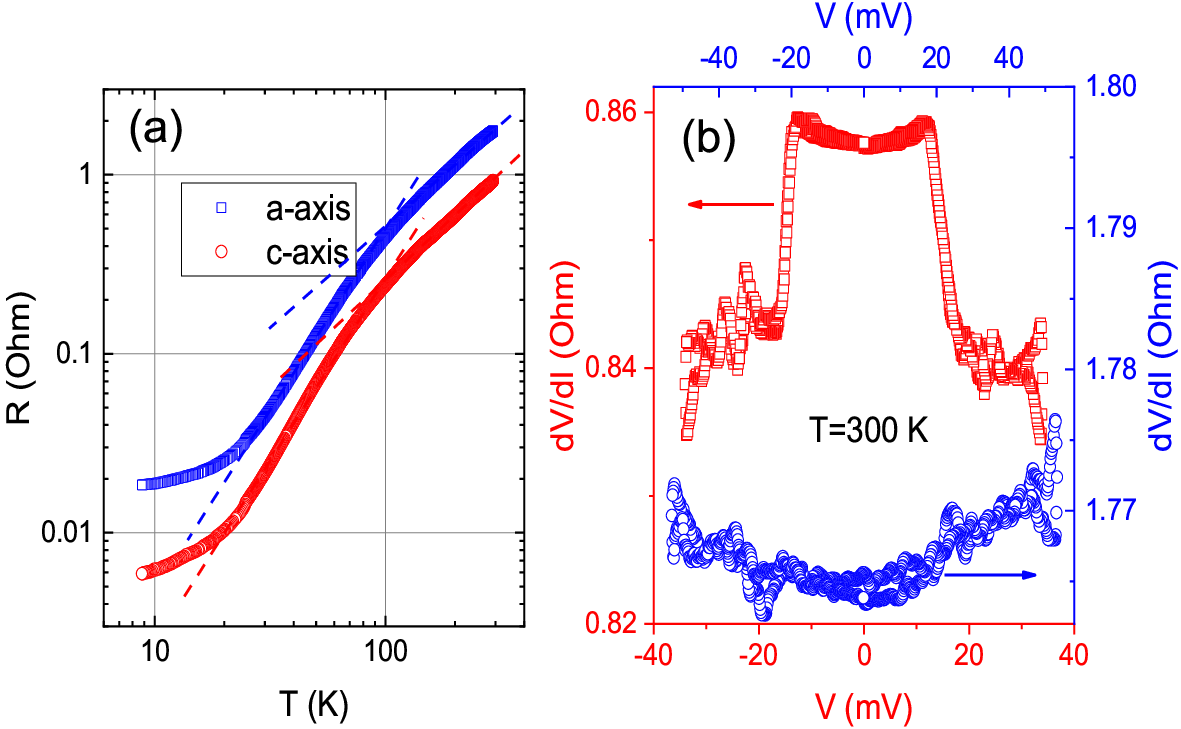}
	\caption{(a) Temperature dependencies of resistance in log-log scale of measured structures in zero magnetic fields when the current applied along the $c$ and $a$-axis. (b) Current-voltage characteristics of measured bridges at $T=300$ K.}
	\label{Fig0}
\end{figure}

From untwinned single crystal of rectangular shape with a thickness 0.6 $\mu$m, we cut bridges along natural orthogonal edges with a width of approximately 50 $\mu$m and length of 3 and 2 mm in the same way as it was done in Ref. \onlinecite{SSC14}. Measurements of electrical resistance have been performed with a conventional 4-probe configuration. For contacts preparation we used gold evaporation and cold soldering by In. Temperature dependencies of resistance, $R(T)$, in the range 300-8 K in log-log scale along the $c$ and $a$ axes for measured structures are shown in Fig. \ref{Fig0}(a). As can be seen, for both structures $R(T)$ follows nearly linear dependence from 300-100 K and a quadratic one from 100-20 K. Resistance residual ratio, $R(300K)/R(8K)$ was 104 and 92 indicating good quality measured structures.

Crystallographic orientation of bridges was determined by measurements of current-voltage characteristics (IVs) at 300 K as shown in Fig. \ref{Fig0}(b). The CDW order appears along the $c$-axis and can only slide along the CDW wave vector \cite{SSC14}. This sliding effect is observable as a sharp decrease of differential resistance when the applied voltage reaches a threshold value for the CDW collective motion. As seen from Fig. \ref{Fig0}(b) this effect exists only for one bridge (red curve) and is absent entirely for the other one. 

The magnetic field was applied sequentially parallel to the main crystallographic axes, $a$, $c$ and $b$ and transverse and longitudinal magnetoresistance was recorded as a function of temperature in the range 4.2-20 K and magnetic field up to 8 T. 

\section{results\label{res}}

\begin{figure*}[tbh]
	\includegraphics[width=0.32\textwidth]{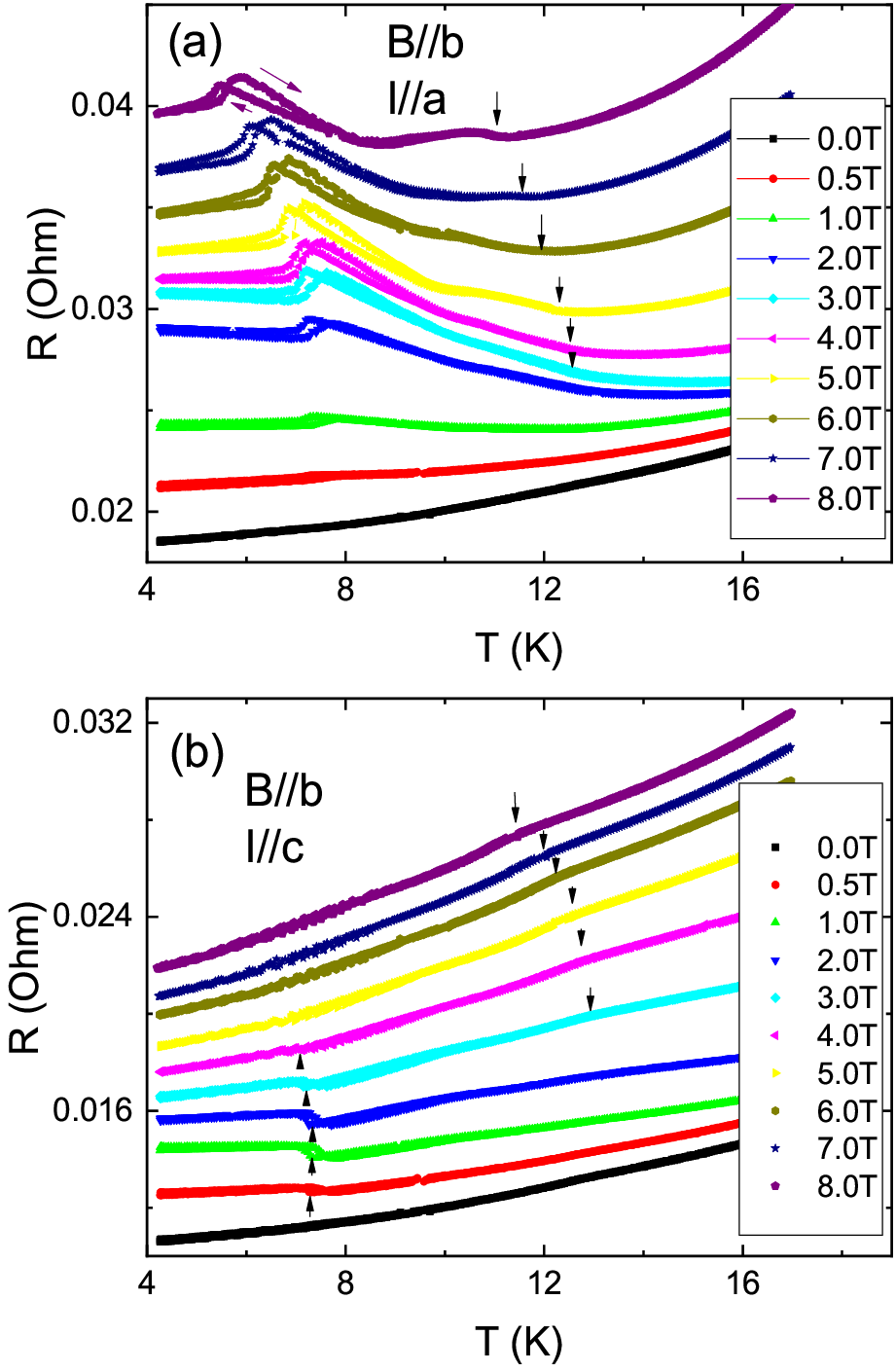}
	\includegraphics[width=0.32\textwidth]{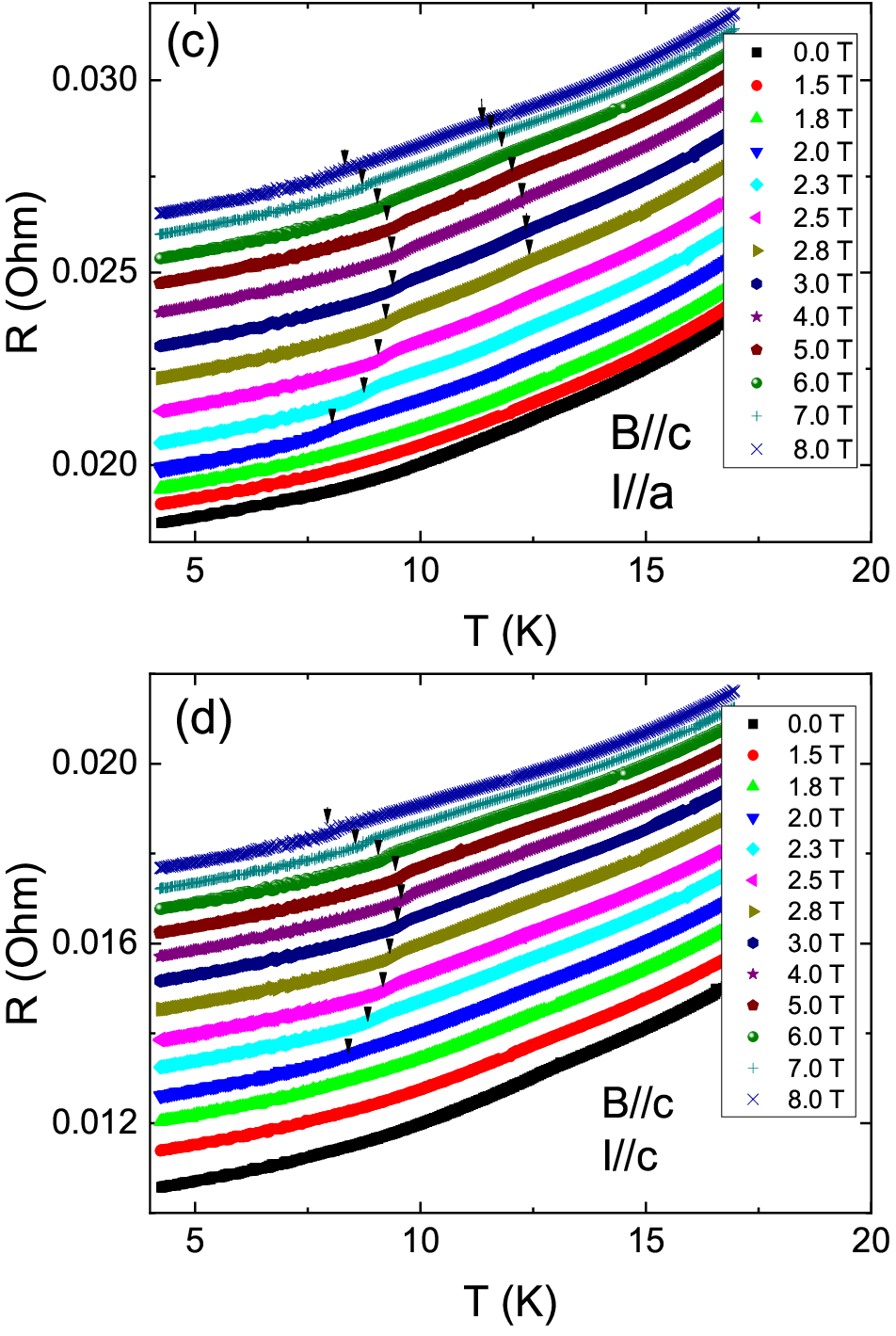}
	\includegraphics[width=0.32\textwidth]{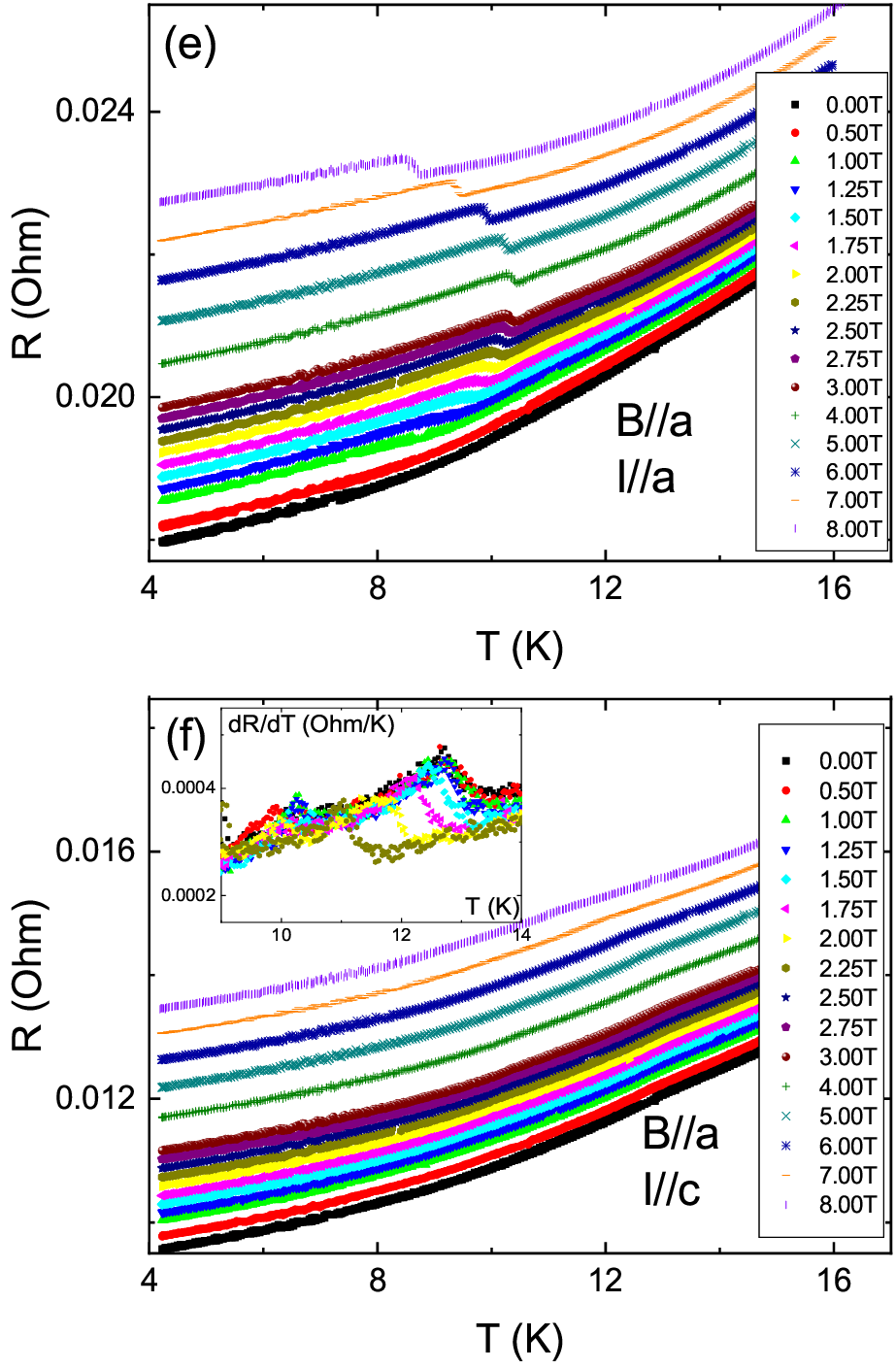}
	\caption{\textbf{Temperature dependencies of resistance in different magnetic fields.} (a) $B\parallel b$ and $I\parallel a$. The curves are shifted for clarity; (b) $B\parallel b$ and $I\parallel c$. The curves are shifted for clarity; (c) $B\parallel c$ and $I\parallel a$; (d) $B\parallel c$ and $I\parallel c$; (e) $B\parallel a$ and $I\parallel a$ and (f) $B\parallel a$ and $I\parallel c$. Inset shows derivative $dR/dT$ as a function of temperature in the range 9-14 K corresponding to the second AFM transition. Positions of singularities corresponding to AFM transitions are indicated by black arrows.}
	\label{Fig1}
\end{figure*}

In Fig. \ref{Fig1} we show the temperature dependencies of resistance for bridges oriented along the $a$ and $c$ axes in magnetic fields $B\leq 8$T oriented along the main crystallographic axes. Remarkably, the experiments at zero magnetic field (black curves in all figures) show no trace of the magnetic transitions observed in $\chi(T)$ and $C(T)$ data. However, application of even a weak magnetic field leads to their appearance indicating additional scattering of normal carriers on magnetic order under such conditions.

Let us first consider the case of a magnetic field oriented along the $b$-axis, $B \perp$ to the basal ($a,c$) plane, (Figs. \ref{Fig1} (a) and (b)). For such a configuration magnetoresistance is large reaching 700-800 \% at $B=8$ T, which makes AFM singularities in $R(T)$ more clear (in all Fig. \ref{Fig1} the curves are vertically shifted for clarity). As can be seen, two AFM transitions are visible for the current applied along the $a$-axis (Fig. \ref{Fig1} (a)) and  $c$-axis (Fig. \ref{Fig1} (b)) with small differences. Thus, for the case $I\parallel a$ the low-temperature AFM transition is observed more clearly at high magnetic field and appears as a decrease of $R$ while for $I\parallel c$ the opposite behavior is found: the effect is detectable at low field and appears as an increase of resistance indicating anisotropic scattering on magnetic order. Note that in both cases we observe a small hysteresis between the increase and decrease of temperature. The high-temperature AFM transition appears similarly in both cases. Both AFM transitions demonstrate a monotonic decrease of transition temperatures with the increase of magnetic field.

In Figs. \ref{Fig1} (c) and (d) we show $R(T)$ curves when the magnetic field is applied in the basal ($a,c$) plane for the case $B\parallel c$ and current configuration $I\parallel a$ and $I\parallel c$ correspondingly. In both cases a smooth but distinguishable decrease of $R$ is observed for the low $T$ AFM transition (indicated by arrows). Note that this transition demonstrates a non monotonic behavior: the AFM transition temperature increases first when $B$ is increased till 4 T and decreases at further increase of magnetic field. For the upper AFM transition we observe a smooth increase of $R$ and the transition temperature decreases monotonically with the increase of $B$. 

A more interesting behavior is observed for the magnetic field parallel to the $a$-axis. $R(T)$ curves for this case are shown in Fig. \ref{Fig1} (e) for current parallel to the $a$-axis and in Fig. \ref{Fig1} (f) for $I\parallel c$-axis. It can be seen that the low-T AFM transition is observed very clearly as a sharp jump of resistance for $I\parallel a$-axis. Under increase of magnetic field the AFM transition temperature demonstrates the same non-monotonic behavior as in the case $B$ parallel to the $c$-axis. The singularity of $R$ corresponding to the high-T transition is absent entirely for this direction of current. Inversely, for $I\parallel c$ the singularity of resistance with the shape of a smooth decrease of $R$ was observed only for the high-T AFM transition while the low-T AFM transition was undetectable. The transition temperature in this case can be determined from the dependence of $dR/dT(T)$ shown in the inset in Fig. \ref{Fig1} (f). 

\begin{figure*}[tbh]
	\includegraphics[width=0.32\textwidth]{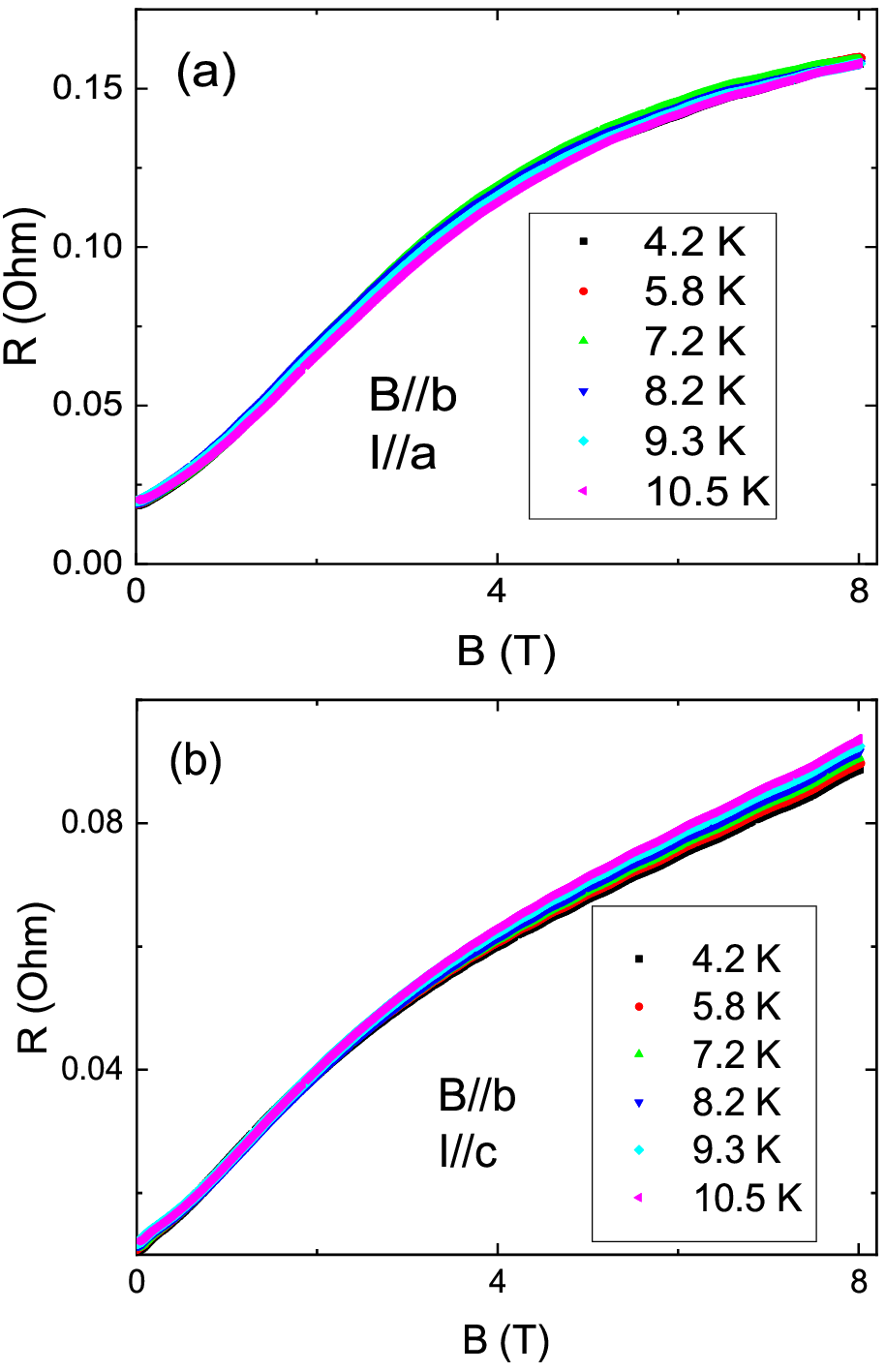}
	\includegraphics[width=0.32\textwidth]{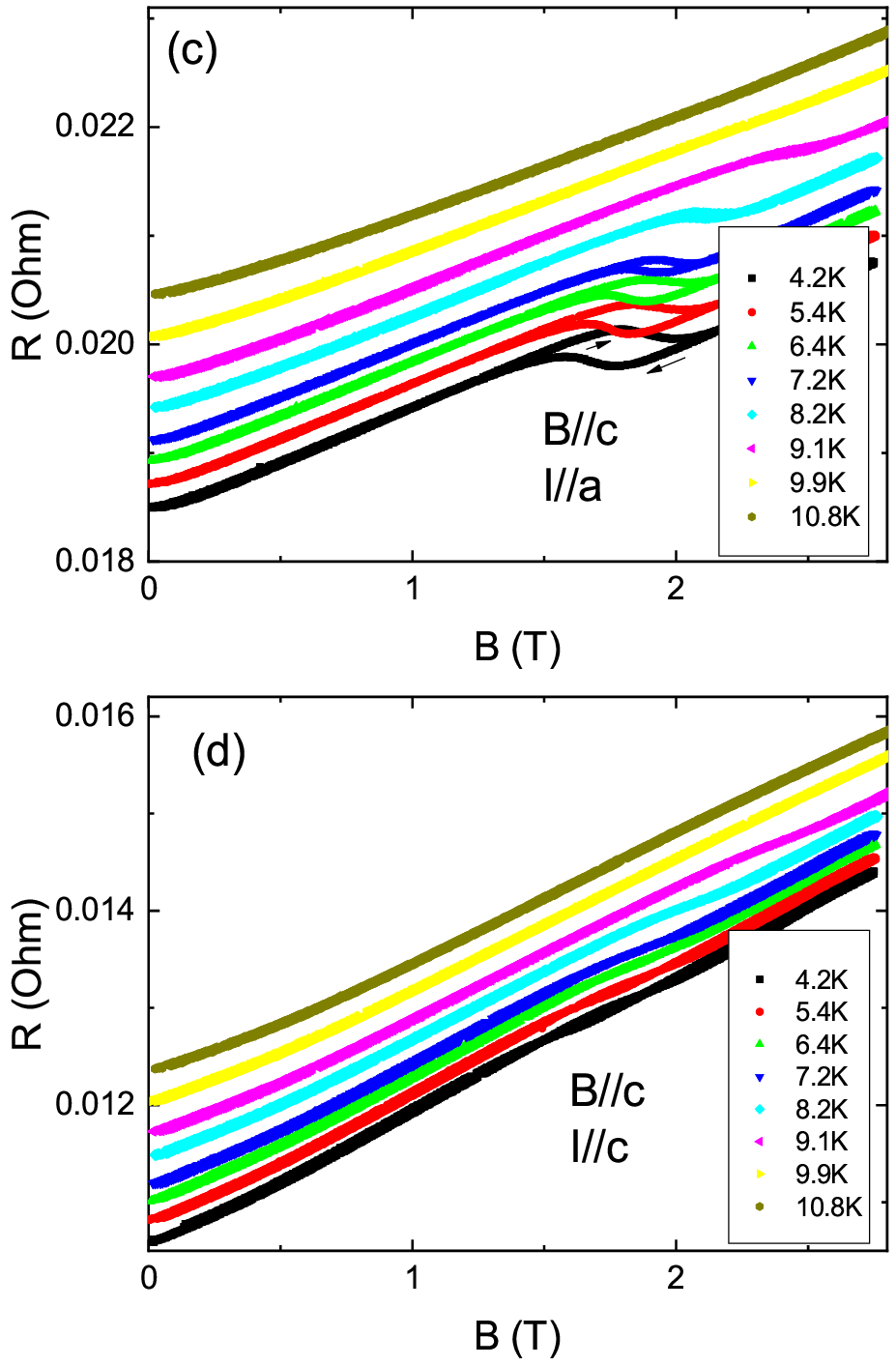}
	\includegraphics[width=0.32\textwidth]{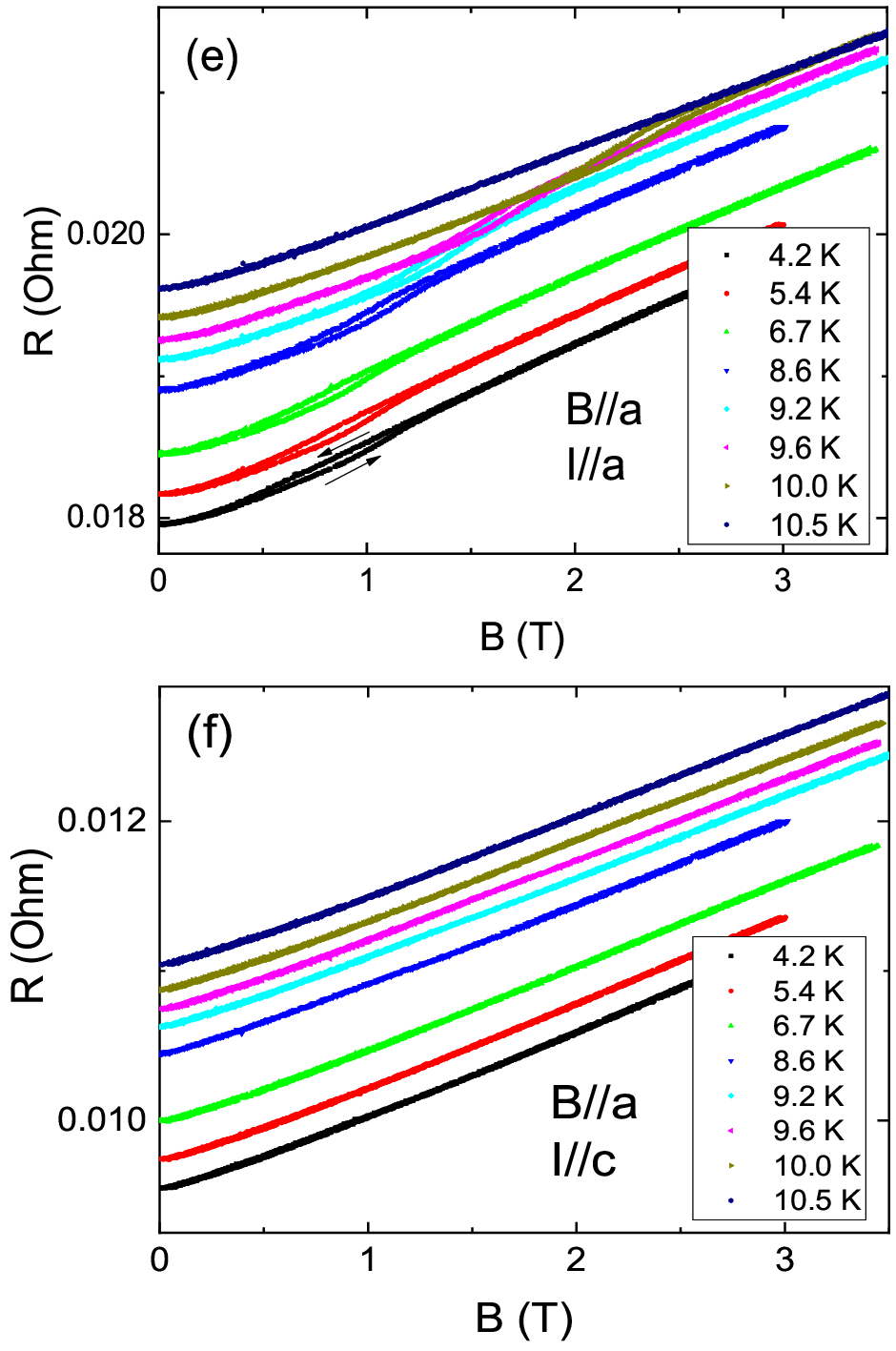}
	\caption{\textbf{Magnetoresistance as a function of magnetic field of different orientation at different temperatures.} (a) $B\parallel b$ and $I\parallel a$; (b) $B\parallel b$ and $I\parallel c$; (c) $B\parallel c$ and $I\parallel a$; (d) $B\parallel c$ and $I\parallel c$; (e) $B\parallel a$ and $I\parallel a$ and (f) $B\parallel a$ and $I\parallel c$.}
	\label{Fig2}
\end{figure*}

Magnetoresistance measurements, $R(B)$, at various magnetic field and current orientations are shown in Fig. \ref{Fig2}. For $B$ perpendicular to the $(a,c)$ plane ($B\parallel b$) magnetoresistance for $I\parallel a$ and $I\parallel c$ are plotted in Fig. \ref{Fig2} (a) and Fig. \ref{Fig2} (b).  For this direction of $B$ magnetoresistance is large and magnetic quantum oscillations (MQO) have been clearly observed for both directions of current in agreement with previously obtained results for GdTe$_3$ \cite{Lei2020,Walmsley2020,EPJB2016}. No effect linked with AFM ordering were observed in this magnetic field configuration.

However, for in-plane oriented magnetic field an effect directly linked with the magnetic ordering state is observed. In Figs. \ref{Fig2} (c) and (d) the $R(B)$ curves at different temperatures are shown for magnetic field oriented along the $c$-axis for $I\parallel a$ and $I\parallel c$ correspondingly. As can be seen, for current along the $a$-axis a sharp decrease of resistance with a pronounced hysteresis loop has been observed at $B=1.8$ T and $T=4.2$ K. With temperature increase the amplitude of this effect decreases and the position of hysteresis loop shifts to higher temperature. For current along the $c$-axis a smooth decrease of resistance with a weak hysteresis of a much less amplitude is also observed in the same magnetic fields region. Figs. \ref{Fig2} (e) and (f) demonstrate magnetoresistance curves at different $T$ for $B\parallel a$ $I\parallel a$ and $I\parallel c$ respectively. As can be seen, for current along the $a$-axis hysteresis effect is also observed but with a much less amplitude compared with $B\parallel c$-axis. This effect is observed in a lower magnetic field region and appears not as a decrease but an increase of resistance. For the case of $I\parallel c$-axis only a small linear magnetoresistance without any peculiarity is observed.

\begin{figure*}[tbh]
	\includegraphics[width=0.3\textwidth]{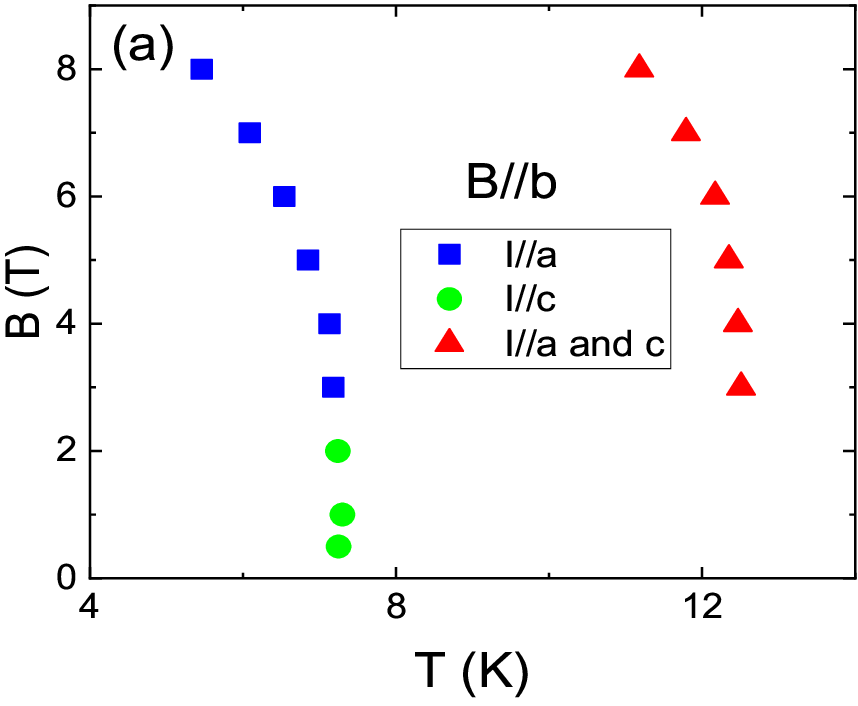}
	\includegraphics[width=0.3\textwidth]{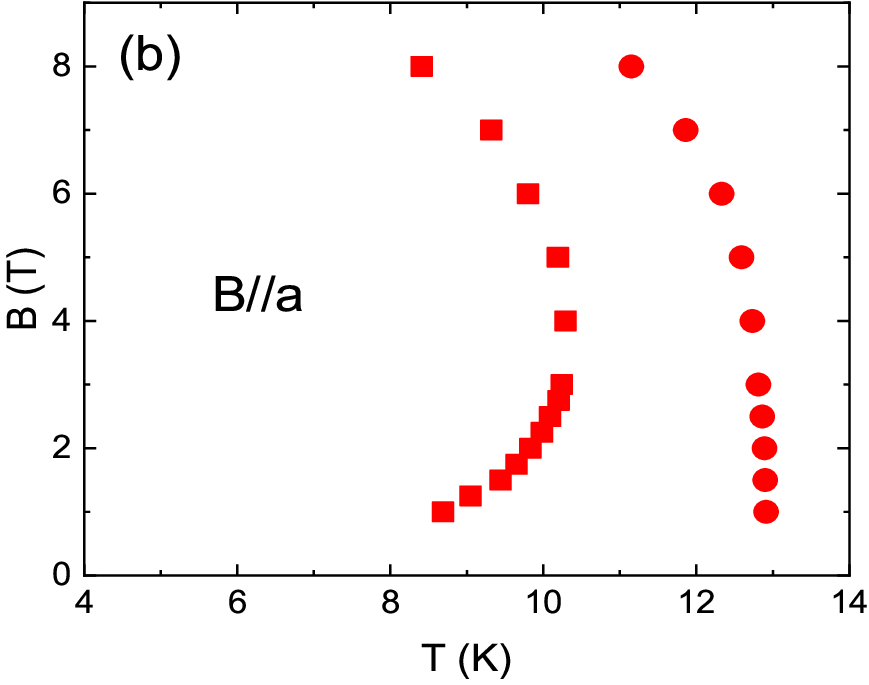}
	\includegraphics[width=0.31\textwidth]{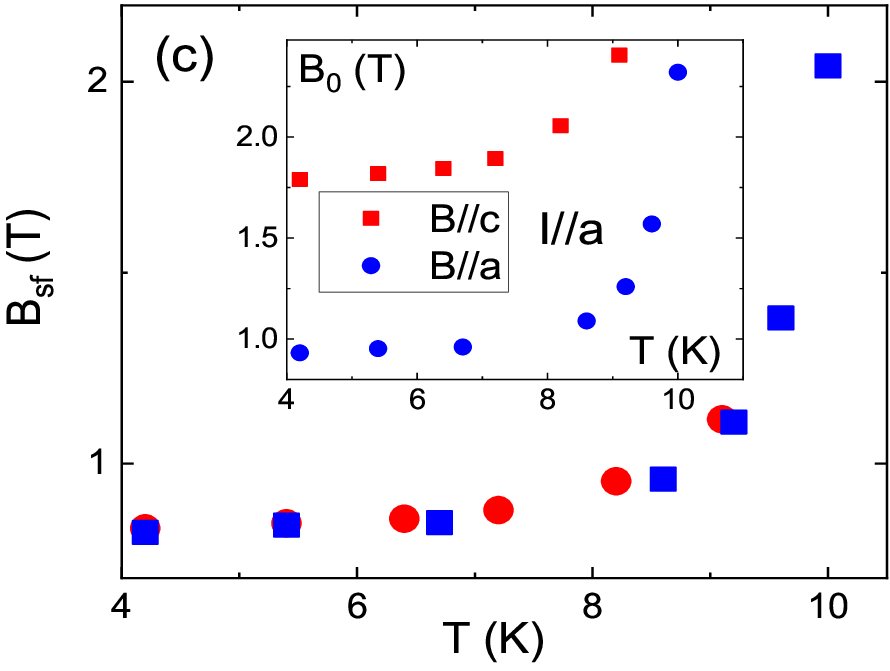}
	\caption{Magnetic phase diagrams GdTe$_3$ for (a) magnetic field $B$ parallel to $b$-axis; (b) magnetic field $B$ applied along the ($a,c$) plane. (c) Dependency of spin flop critical field, $B_{sf}$, on temperature. Inset shows temperature dependencies of critical field, $B_0$, for $B$ oriented along $c$-axis (red squares) and $a$-axis (blue circles).}
	\label{Fig3}  
\end{figure*}

Previously at the same scale of magnetic field a hysteresis effect in GdTe$_3$ was observed \cite{Iyeiri03} in the magnetization curve when changing its slope at 2.5 K for in-plane magnetic field, suggesting a spin-flop transition. Unfortunately, the exact direction of magnetic field relative to crystallographic axes was not determined. In the spin-flop transition scenario once the magnetic field oriented along the easy axis hits the critical threshold, it becomes energetically favorable for the spins to "flop". They rotate away from the easy axis and orient themselves nearly perpendicular to the magnetic field. Observation in our experiments of a hysteresis loop indicates directly a first-order phase transition that is a property of spin flop reorientation transition in easy-axis antiferromagnets. Note that the effect is strongly anisotropic with respect to the direction of current indicating an anisotropic normal carrier scattering on magnetic order.

\section{discussion and conclusion\label{res}}

The data obtained allow us to draw magnetic phase diagrams of GdTe$_3$ from electron transport measurements only. In Fig. \ref{Fig3} (a) the $B-T$ phase diagram is drawn with the magnetic field applied along the $b$-axis (perpendicular to the basal plane). We identified two AFM transitions with extrapolation at zero magnetic field at $T_{N1}=12.5$ K and $T_1=7$ K, the transition temperature of which decreases monotonically with increase of magnetic field. Our results agree with measurements of magnetic susceptibility at the same magnetic field orientation which have revealed the three magnetically ordered phases at $T_{N1}=12$ K, $T_{N2}=10.5$ K and $T_1=7$ K \cite{Lei2021}. Note that in this $B\parallel b$ configuration we have not detected any peculiarities in $R(T)$ curves which can correspond to magnetic ordering observed at $T_{N2}=10.5$ K at zero field. Probably, this transition does not affect normal metal scattering. Hysteresis in the low temperature phase transition indicates its first order nature.

In Fig. \ref{Fig3} (b) we show the $B-T$ phase diagram for magnetic field oriented along the $a$-axis. In this figure we only show data for $B\parallel a$-axis configuration because, although for ($B\parallel c$-axis) the same behavior is observed, the AFM peculiarities are much weaker (Fig. \ref{Fig1} (c) and (d)). One can conclude that the positions of AFM peculiarities in $R(T)$ are independent of magnetic field and current directions in the case $B\perp b$-axis. The transition temperature of the upper AFM order demonstrates the same behavior as in the case $B\parallel b$ with extrapolation at zero field at $T_{N1}=12.5$ K and a monotonic decrease $T_{N1}$ with increase of magnetic field. Our data qualitatively coincide with previous results \cite{Guo2021} (with a slight difference in the absolute value of the transition temperature). Surprisingly the transition temperature of the low-T AFM exhibits a non-monotonic dependence on magnetic field increasing at low field till a maximum value around $T=10$ K and decreasing at higher magnetic field. But no trace of the $T_{N2}$ AFM transition is detectable in our transport measurements.

Fig. \ref{Fig3} (c) shows the temperature dependence of spin-flop threshold magnetic field, $B_{sf}$ extracted from Figs. \ref{Fig2} (c) and (e). To get this value we assumed that when the magnetic field is applied in the ($a,c$)-plane, only the projection of applied field on the easy axis provides the spin-flop effect. This assumption explains the observed difference in magnitudes of threshold magnetic fields, $B_0$, when the magnetic field is oriented along the $a$-axis (Fig. \ref{Fig2} (e)) and $c$-axis (Fig. \ref{Fig2} (c)) such as $B_0^{a}\sin\alpha=B_0^{c}\cos\alpha$ with $\alpha$ being the angle between easy magnetisation axis and $c$-axis. $B_0^{c}/B_0^{a}=\tan\alpha$ must be a constant as experimentally found. We get $\tan\alpha=1.90\pm 0.3$ that corresponds to $\alpha\approx 62^\circ$. Thus, it follows from our data that the easy axis is oriented at 62 degrees relative to crystallographic $c$-axis. In Fig. \ref{Fig3} (c) it is plotted $B_{sf}=B_0^{a}\sin\alpha$ as red circles and $B_{sf}=B_0^{c}\cos\alpha$ as blue squares and corresponding dependencies $B_0^{c}(T)$ and $B_0^{a}(T)$ in inset. In both cases, the position of peculiarities are determined for the regime of increasing magnetic field.

We raise now the question, why the electric resistance measurements give the AFM phase diagram, and especially, why peculiarities in resistance measurements, inexistent at zero magnetic field, only arise under application of magnetic field? There are at least two opposite and competing effects of AFM phase transition on resistivity. The observed complicated dependence of magnetoresistance at AFM phase transitions on the directions of magnetic field and current can be qualitatively understood as a result of this competition. 

First, the formation of long-range magnetic order at $T_{1}\approx 7$\thinspace K suppresses the AFM fluctuations. The AFM fluctuations mostly have the wave vector close to $\boldsymbol{Q}_{AFM}$, the wave vector of long-range AFM order at lower $T$. Hence, the additional scattering of conducting electrons by AFM fluctuations mostly affect the current along $\boldsymbol{Q}_{AFM}$, i.e. $I\parallel a$. The suppression of AFM fluctuations at $T_{1}$ reduces resistivity at temperature $T<T_{1}\approx7$\thinspace K slightly below the second AFM phase transition $T_{AFM}$. Therefore, the temperature of resistivity drop may be slightly below $T_{AFM}$ detected from susceptibility measurements. The enhancement of electron scattering by AFM fluctuations extends to a large temperature range $T_{1}<T\lesssim
T_{N1}\approx 12.5\,$K and even above $T_{N1}$, where it decreases smoothly with the increase of $T$. Therefore the onset of this scattering mechanism at $T>T_{N1}$ is not sharp and can hardly be detected. 

The second effect of AFM phase transition is related to the electron scattering by AFM wave vector $\boldsymbol{Q}_{AFM}=\left( \pi/a,0\right)\perp\boldsymbol{Q}_{CDW}$. Unfortunately, we do not know the value of interlayer component of $\boldsymbol{Q}_{AFM}$, probably, it is either $0$ or $\pi /b$. The scattering by AFM order parameter changes the electron dispersion and mainly affects the current along $\boldsymbol{Q}_{AFM}$, i.e. along the $a$-axis rather than along $ \boldsymbol{Q}_{CDW}||c$-axis, because the electron momentum relaxation increases only along $\boldsymbol{Q}_{AFM}||a^{\ast }$. Therefore, its effect is much stronger in Figs. 2a and 2e than in Figs. 2b and 2f. This scattering mechanism works only in the special hot spots on the 2D Fermi surface (FS), or on hot lines almost $||b$ in 3D FS, which are connected by $\boldsymbol{Q}_{AFM}$. Hence, only when electrons come to these hot spots or lines, they get additional scattering by $\boldsymbol{Q}_{AFM}$. This results to a strong dependence of this mechanism on the strength and direction of magnetic field $\boldsymbol{B}$. 

The strongest effect of hot spots is expected for $\boldsymbol{B}||\boldsymbol{b}$, when most electrons pass through the hot spots. For $\boldsymbol{B}||\boldsymbol{b}$ the electron momentum rapidly moves along the Fermi surface due to the large Lorentz force $F=\left[ \boldsymbol{v}\times \boldsymbol{B}\right] e/c$. The latter is large because the electron velocity $\boldsymbol{v}\perp \boldsymbol{B}$, being mostly directed along the conducting plane. Among the in-plane directions of magnetic field, the
scattering rate in hot spots is stronger at $\boldsymbol{B}||\boldsymbol{a}$ than at $\boldsymbol{B}||\boldsymbol{c}$, because close to hot spots the electron velocity $\boldsymbol{v}||\boldsymbol{c}$, because at the boundary $k_{a}=\pi /\left( 2a\right) \boldsymbol{\ }$of magnetic Brillouin zone the normal-to-boundary velocity component vanishes. Therefore, one observes a jump of resistance in Fig. 2e when the temperature is reduced below $T_{1}$. 

Additionally, the symmetry between two main in-plane directions of magnetic field is lifted by the direction of the easy magnetization axis, giving the direction of AFM order parameter $\mathbf{\Delta}_{\text{AF}}$, because the electron dispersion and symmetry in the presence of commensurate AFM order depends on mutual orientation of $\mathbf{\Delta }_{\text{AF}}$\ and $\boldsymbol{B}$ \cite{RamazashviliPRB2009,Ramazashvili2021}. 
At high field, $B>B_{sf}$, $\mathbf{\Delta }_{\text{AF}}\perp \boldsymbol{B}$ for any direction of $\boldsymbol{B}$, but even in this case the strength of AFM order and of its fluctuations depends on mutual orientation of $\boldsymbol{B}$ and of the easy magnetization axis. 

The higher the magnetic field, the larger the Lorenz force and the electron scattering rate in hot spots. In clean samples, when $\omega _{c}\tau >1$, most electrons have time to reach the hot spot before they scatter by crystal defects. This results to linear magnetoresistance, similar to the effect of CDW hot spots \cite{Falicov1972,Sinchenko2017}, in contrast to usual quadratic magnetoresistance with saturation. This linear magnetoresistance contribution is observed in all geometries, as shown in Fig. 3.

Interplay between magnetism and CDWs has also been investigated in other RTe$_3$ compounds. Thus in TbTe$_3$, AFM transitions occur with $q_{m1}=(1/2,1/2,0)$ below 6.6 K, and $q_{m2}=(0,0,1/2)$ below 5.6 K with the additional incommensurate AFM below $T_{N3}=5.4$ K with the propagation vector being $q_{AFM}-q_{CDW}$ \cite{Chillal2020}. In DyTe$_3$ \cite{Alekseev2024} similarly several types of magnetic peaks, commensurate with $q_{AFM}=0.5c^*$ and incommensurate one with $q_{IC}=0.27c^*=0.5-q_{CDW}$ with $q_{CDW}=2/7c^*=0.29c^*$ have been observed. The CDW satellites in CeTe$_3$ appear \cite{Okuma2026} at $q_{CDW}=0.28c^*$ but the incommensurate magnetic reflections have a transverse component such as $q_m=0.17a^*\pm 0.31c^*$ with the $c^*$ component equal to 0.5 minus the CDW component. Note that there is no coupling between CDW and AFM in HoTe$_3$ \cite{Yun2026}.

Charge order and magnetic orders in GdTe$_3$ were recently studied via spin polarized scanning tunneling microscopy \cite{Raghavan2024}. It was shown that an intermediate phase (a spin density wave in Ref. \onlinecite{Raghavan2024}) sets up in combination of a $b$-axis AFM order with $q_{AFM}=(\pm\pi/a,0)$ (we swap the space group Bmcm used in reference to Cmcm in present paper) and the CDW order $q_{CDW}=(0,2/7c^*)$. This intermediate phase was shown to disappear below 7-8 K with the $b$-axis AFM present between 7.5 and 12 K rotating to the ($a,c$) plane below 7 K where it becomes striped AFM observed. 

The distortion vectors of two upper AFM orders in GdTe$_3$ are unknown. From our measurements one can say that the upper AFM transition with its phase diagram in both magnetic field directions parallel and perpendicular to the ($a,c$) basal plane is robust and very probably does not interact with the CDW. In Ref. \onlinecite{Raghavan2024} it was only considered a single AFM presumably $T_{N1}$ one. Taking into account the specific heat measurements under magnetic field applied in the basal plane \cite{Guo2021} one can see that the anomaly in $C_p$ around the $T_1$ transition is very weak which makes that this transition was not detected in some works. The temperature dependence of the $T_{N2}$ transition decreases slightly till merging around 2-3 T with the anomaly corresponding to the $T_1$ transition with the consequence of a huge peak in $C_p$. 

We have measured a spin-flop transition down to 4 K. The temperature of the spin-flop increases sharply in the temperature range 8-10 K where it "crosses" the re-entrance temperature dependence of peculiarities in the $R(T)$ at fixed field with the possibility of a triple point. Our measurements, however, are too macroscopic for being more precise and further investigations are necessary. 

Our results demonstrate the use of low-temperature electronic transport measurements to construct a $B-T$ magnetic phase diagram, and they further develop this methodology, which is applicable to a wide range of compounds. This approach is particularly advantageous for GdTe$_3$, a material for which neutron scattering experiments are extremely challenging due to the high neutron absorption by Gd atoms. Magnetotransport measurements are simpler to implement than other techniques under high pressure, making them useful for constructing $B-T-P$ phase diagrams. Although a rigorous theoretical foundation is still lacking, magnetoresistance features associated with magnetic phase transitions had already been reported and used to study the magnetic phase diagram in other materials, such as Ce$_3$Pd$_6$Sb$_5$ \cite{Xie2024}, CeBi \cite{Kuthanazhi2022}, HoPtSn \cite{Agarwal2026}, NdAuAl$_4$Ge$_2$ \cite{Cong2023}, and others. 

In conclusion, our measurements emphasize the specific part in the phase diagram of GdTe$_3$ between 8-10 K and with the magnetic field 1.5-3.0 T with interaction between a CDW with a possible orbital structure, antiferromagnetic structures occurring at $T_{N2}$ and $T_1$ at zero field and spin-flop transition in the basal plane.

\begin{acknowledgments}
	The work is partially supported by the Russian Science Foundation Project No. 26-12-00345.
\end{acknowledgments}

\end{document}